\documentclass[journal=jacsat,manuscript=article]{achemso}
\usepackage[version=3]{mhchem} 

\usepackage{kotex}
\usepackage{multirow}
\usepackage{adjustbox}
\usepackage{color}
\usepackage{xcolor}
\usepackage{hyperref}

\newcounter{daggerfootnote}

\author{Seongok Ryu}
\affiliation[GALUX]{GALUX inc., Seoul, 08738, Republic of Korea}
\email{seongokryu@galux.co.kr}

\author{Sumin Lee}
\affiliation[SNU]{Department of Chemistry, Seoul National University, Seoul, 08826, Republic of Korea}

\title[An \textsf{achemso} demo]
  {Accurate, reliable and interpretable solubility prediction of druglike molecules with attention pooling and Bayesian learning}

\abbreviations{Solubility, Drug discovery, Machine Learning, Deep Learning, Graph Neural Networks, Bayesian Learning, Attention}

\begin{document}

\begin{abstract}
In drug discovery, aqueous solubility is an important pharmacokinetic property which affects absorption and assay availability of drug. 
Thus, \textit{in silico} prediction of solubility has been studied for its utility in virtual screening and lead optimization.
Recently, machine learning (ML) methods using experimental data has been popular because physics-based methods like quantum mechanics and molecular dynamics are not suitable for high-throughput tasks due to its computational costs. 
However, ML method can exhibit over-fitting problem in a data-deficient condition, and this is the case for most chemical property datasets. 
In addition, ML methods are regarded as a black box function in that it is difficult to interpret contribution of hidden features to outputs, hindering analysis and modification of structure-activity relationship.
To deal with mentioned issues, we developed Bayesian graph neural networks (GNNs) with the self-attention readout layer. 
Unlike most GNNs using self-attention in node updates, self-attention applied at readout layer enabled a model to improve prediction performance as well as to identify atom-wise importance, which can help lead optimization as exemplified for three FDA-approved drugs.
Also, Bayesian inference enables us to separate more or less accurate results according to uncertainty in solubility prediction task
We expect that our accurate, reliable and interpretable model can be used for more careful decision-making and various applications in the development of drugs.
\end{abstract}

\section{Introduction}
\begin{figure}
    \centering
    \includegraphics[width=0.48\textwidth,trim={0cm 0 0cm 0},clip]{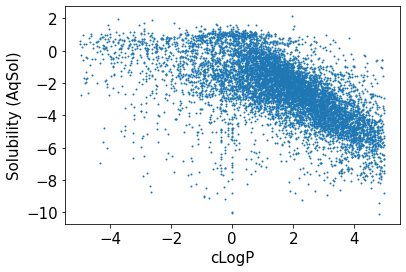}
    \caption{Scatter plot between cLogP values obtained by RDKit and solubility values provided by AqSolDB\cite{sorkun2019aqsoldb}.}
    \label{fig:solubility_clogp}
\end{figure}

Aqueous solubility (hereinafter referred to as "solubility") is one of the most necessary pharmacokinetic properties in drug discovery, since it affects absorption, bioavailability and toxicity of drug candidates.
Predicting solubility is also essential for early-stage drug discovery, as insoluble compounds may not be available for biochemical assays. 
For these two reasons, \textit{in silico} prediction of solubility has been heavily investigated. 
An alternative approach is to use freely available cheminformatics software such as RDKit\cite{landrum2013rdkit} to obtain water-octanol partition coefficient (cLogP).
However, as shown in Figure \ref{fig:solubility_clogp}, cLogP is not a good alternative to experimentally-determined solubility. 

More sophisticated approaches, like other chemical property prediction problems, can be categorized into three: quantum mechanics (QM), molecular dynamics (MD), and data-driven methods. 
COSMO-RSol calculates solubility based on the free energy change of solid to liquid state using a quantitative structure-property relationship (QSPR) model to estimate free energy change of fusion and QM calculation to estimate the difference between chemical potential of pure and solution states.\cite{klamt2002prediction}
MD predicts solubility through a two-phase crystal-liquid system simulation.
The effect of transition can be calculated by integrating simulations performed on a crystal or a liquid system independently or by metadynamics of a coexisting system directly.\cite{bjelobrk2021solubility}
However, due to heavy computational cost, both QM and MD approaches may be insuitable for virtual screening. 
On the other hand, data-driven methods, which are mostly based on machine learning (ML), learn the relationship between molecular features and target properties given training data samples.
However, as these models require training data samples, inferior size and quality of available data have hindered their implementation to virtual screening.\cite{sorkun2019aqsoldb}

Lessons learnt from the previous solubility prediction challenges, recent works has begun to focus on data curation and feature engineering.\cite{sorkun2019aqsoldb, sorkun2021pushing, lovric2021machine}
In the first challenge sponsored by Pfizer and hosted by Goodman group,\cite{llinas2008solubility} most submitted models showed lower than 60\% accuracy,\cite{hopfinger2009findings} which was criticized to be due to improper algorithms or molecular features rather than the noisy data (i.e., participants were provided with only 100 training data samples).\cite{palmer2014experimental}
However, recent research demonstrates that simple algorithms such as support vector machine and partial least squares with well-curated data showed comparable performance.\cite{lovric2021machine} 
Results from the second solubility challenge also back up the importance of data quality and informative descriptors.\cite{llinas2020findings} 
There was no significant difference between models trained with thousands of samples and only few hundreds samples because both datasets were not able to sufficiently account for chemical space of the drug-like molecules.
In addition, state-of-the-art ML algorithms, such as graph neural networks and XGBoost, exhibited similar or even inferior performance to multiple linear regression.


Previous studies have mostly focused on improving prediction accuracy, but data-driven models show deteriorated performance on predicting unseen molecules as exemplified in the previous challenges.
Therefore, we would like to argue that it is necessary to assess models in terms of not only accuracy but also predictive uncertainty and interpretability for real task.
For example, if a given query is an out-of-distribution sample, a model should exhibit high predictive uncertainty, which can help users to agree or disagree the prediction results.\cite{ryu2019bayesian, schwaller2019molecular, soleimany2021evidential, graff2021accelerating}
In addition, interpretability, such as contribution of molecular substructures to predictive solubility, enables users to infer how improve solubility by modifying substructures.\cite{harren2022interpretation}

In this work, we present a solubility prediction model with accurate, reliable and interpretable results. 
In contrast to QSPR models based on molecular descriptors and old-fashioned ML models (e.g. support vector machine and random forest), we utilized graph neural networks in order to learn features from graph representation of molecular structures. 
Also, we adopt self-attention\cite{vaswani2017attention} in the readout function, which aggregates atom features and produces a graph feature per single molecule.
Our designed readout not only improves prediction performance but also enables us to interpret the attention weight as importance value.
Lastly, we confirm that using Bayesian inference results in improved prediction performance than non-Bayesian approach and uncertainty quantification, which can be utilized to prioritize more accurate results based on predictive uncertainty.

\section{Methods and Implementations}
\subsection{Dataset and data splitting}
We used the AqSolDB dataset \cite{sorkun2019aqsoldb} provided by Therapeutic Data Commons \cite{huang2021therapeutics} to develop our neural networks.
This dataset consists of 9,982 molecules, where the structures are given in simplified molecular-input line-entry system (SMILES) format and the labels are annotated with continuous solubility indicator value. 
As described in \citeauthor{huang2021therapeutics}\cite{huang2021therapeutics}, molecules can be categorized by the indicator value ($y_{true}$) as follows:
\begin{itemize}
    \item $y_{true} > 0.0$ : highly soluble 
    \item $ -2.0 < y_{true} \leq 0.0 $ : soluble
    \item $ -4.0 < y_{true} \leq -2.0 $ : partially soluble
    \item $y_{true} < -4.0$ : insoluble
\end{itemize}

We developed our models with those SMILES and label pairs with the four-fold scaffold splitting\cite{wu2018moleculenet} of entire set into 70\%/10\%/20\% for train/validation/test sets.
Both classification and regression models were obtained with and without converting the indicator values to the categorical labels, respectively.

\subsection{Input featurization and Graph neural networks}
Our baseline model framework is graph neural network (GNN) which utilizes the graph representation of molecular structure as their inputs.
To generate inputs for our model, we featurize molecular graphs $G(V,E)$, where $V$ is the set of node features $\{ h_i^{(0)} \}$, $E$ is the set of edge features ${\{ e^{(0)}_{ij}} \}$ and $i$ and $j$ are node indices, with RDKit\cite{landrum2013rdkit}.
We assign initial features to each nodes by using the RDKit functions as follows:
\begin{itemize}
    \item One-hot encoding of atom types: atom.GetSymbol()
    \item One-hot encoding of atom degree: atom.GetDegree()
    \item One-hot encoding of number of hydrogens: atom.GetTotalNumHs()
    \item One-hot encoding of implicit valence number: atom.ImplicitValence()
    \item Indicator value for whether given atom is aromatic: atom.GetIsAromatic()
\end{itemize}
Also, initial edge features are assigned as follows:
\begin{itemize}
    \item One-hot encoding of atom bond types: see our code released in github.
    \item Indicator value for whether bond is conjugated: bond.GetIsConjugated()
    \item Indicator value for whether bond is in ring: bond.IsInRing()
\end{itemize}
\citeauthor{hwang2020comprehensive} shows that descriptors, which can appropriately describe atom and bond characters, significantly affects prediction performance of GNNs, 
and thus we follow the input featurization scheme of the previous work.\cite{hwang2020comprehensive}

Using the featurized input molecular graphs, the $l$-th message passing layer updates the $i$-th node feature $h_{i}^{(l-1)}$ to the output $h_{i}^{(l)}$ by aggregating the adjacent node and edge features:
\begin{equation}
    h_i^{(l)} = \text{NodeUpdate} (h^{(l-1)}_{i}, \{h^{(l-1)}_{j}\}_{j \in \mathcal{N}_{i}}, \{e^{(l-1)}_{ij}\}_{j \in \mathcal{N}_{i}}), 
\end{equation}
where $\mathcal{N}_{i}$ stands for the indices of the neighbor nodes of the $i$-th node. 

After passing the $L$-stack of message passing layers, the node features are aggregated and summarized to a single vector $z_{G}$, so called the graph feature, by the readout layer:
\begin{equation}
    z_{G} = \text{Readout}(\{h_i^{(L)}\}).
\end{equation}

Then, the final prediction layer converts the graph feature to a predictive value:
\begin{equation}
    y_{pred} = \text{Linear}(z_{G}).
\end{equation}

\subsection{Node updates}
We implemented three different GNN node updates and then chose the best model based on the results from the first validation experiment, as shown in Figure \ref{fig:RMSE_R2}.
Graph convolutional network (GCN)\cite{duvenaud2015convolutional, kipf2016semi} is the simplest node update among the three implementations.
The node update equation of GCN can be described as follows:
\begin{equation}
        h_i^{(l)} = \text{LN}(h_i^{(l-1)} + \sigma(W^{(l)}\sum_{j \in \mathcal{N}_{i}} h^{(l-1)}_{j})),
\end{equation}
where $\text{LN}$ stands for layer normalization\cite{ba2016layer} and $\sigma(\cdot)$ is non-linear activation. 

Graph isomorphism network (GIN)\cite{xu2018powerful} can be thought as more sophisticated version of GCN, which replaces the linear layer of GCN to the multi-layer perceptron (MLP):
\begin{equation}
            h_i^{(l)} = \text{LN}(h_i^{(l-1)} + \text{MLP}^{(l)} (\sum_{j \in \mathcal{N}_{i}} h^{(l-1)}_{j})).
\end{equation}
where $\text{MLP}$ stands for
\begin{equation}
    \text{MLP}^{(l)}(x) = W_2^{(l)}\sigma(W_1^{(l)}x + b_1^{(l)}) + b_2^{(l)}
\end{equation}
Furthermore, to utilize bond information, we include edge features to the GIN layer's aggregation of node features:
\begin{equation}
    h_i^{(l)} = \text{LN}(h_i^{(l-1)} + \text{MLP}^{(l)} (\sum_{j \in \mathcal{N}_{i}} h^{(l-1)}_{j} + e_{ij}^{(l-1)})).
\end{equation}
We expect that the GIN shows better results than the GCN as the former transforms both node and edge features of adjacent nodes. 

Lastly, we implemented a modified version of graph attention network\cite{velivckovic2017graph}, which adopts self-attention in node updates. 
Our modified version, namely graph transformer (GT), incorporates edge features in both computation of attention weights and aggregation of adjacent node features:
\begin{equation}
    h_i^{(l)} = \text{LN}(h_i^{(l-1)} + \text{MLP}^{(l)} (\hat{h}_{i}^{(l)})),
\end{equation}
where the intermediate node feature $\hat{h}_{i}^{(l)}$ is given by
\begin{equation}
    \hat{h}_{i}^{(l)} = \text{LN}(W_{6}^{(l)}h^{(l-1)}_{i} + \sum_{j \in \mathcal{N}_{i}} w_{ij}^{(l)} (W_{5}^{l} h^{(l-1)}_{j} + W_{4}^{(l)} e_{ij}^{(l-1)}) ).
\end{equation}
The attention coefficient $w_{ij}^{(l)}$ between two adjacent nodes $i$ and $j$ is computed by softmax activation of scaled-dot products between the source node feature and the adjacent node and edge features.:
\begin{equation}
    w_{ij}^{(l)} = \text{softmax}(\frac{1}{\sqrt{d}} (W_{3}^{(l)} h_{i}^{(l-1)}) (W_{2}^{(l)} h_{j}^{(l-1)} + W_{1}^{(l)} e_{ij}^{(l-1)})^{T} ).
\end{equation}
We note that our implementation of GIN and GT is inspired by that introduced in PyTorch Geometric\cite{fey2019fast} implementations.

\subsection{Readout with self-attention}
In most GNN implementations, the readout (or sometimes referred to as pooling) step adopts mean and sum operations to aggregate node features that are updated by a stack of message passing layers. 
Both operations do not assign different importance to each node, but aggregates node features with equal weights.

On the other hand, we hypothesize that assigning different atom-wise importance to each node results in more accurate prediction:
\begin{equation}
    z_{G} = \sum_{v} \alpha_v h_v^{(L)},
\end{equation}
where $\alpha_v$ is the attention weight given to the $v$-th node, since attention weights can be considered as the importance of objects -- in our case node (atom) features.
We expect that adopting attention weights is not only improves prediction performance but also helps model interpretability, as is validated by computer vision and language understanding works\cite{dosovitskiy2020image, vaswani2017attention}. 

Therefore, we utilize pooling by multi-head attention (PMA) firstly introduced in Set Transformer\cite{lee2019set}, 
where attention weight $\alpha_v$ is given by 
\begin{equation}
    \alpha_v = \text{softmax}(\frac{1}{\sqrt{d}} \textbf{1} (W_{\text{pma}} h_{v}^{(L)})^{T} ).
\end{equation}
Although PMA implemented in Set Trasanformer uses a  randomized seed vector as a query vector, 
we found that a fixed vector \textbf{1} initialized with one shows better performance in our solubility prediction task.

In our experiments, we compare mean and PMA readouts as shown in Figure \ref{fig:RMSE_R2}.
As described in \citeauthor{xu2018powerful}, mean and sum readouts are appropriate to summarize node statistics in terms of fraction and exact number, respectively.\cite{xu2018powerful} 
\textcolor{blue}{
In solubility prediction, using ratio between hydrophilic and hydrophobic groups is more appropriate than exact number of those groups to summarize node statistics.
}
In this work, we conjecture and experimentally confirm that sum readout leads to unstable training and mean readout is appropriate.

\subsection{Bayesian learning for uncertainty quantification and reliable prediction}
Our model quantifies uncertainty and calibrates predictive label with Bayesian learning and inference.
As shown in \citeauthor{ryu2019bayesian}, uncertainty quantification enables detecting erroneously labeled samples in a dataset based on the data-driven (aleatoric) uncertainty.
This offers that uncertainty can be utilized to prioritize predictions expected to be more accurate.
Following a benchmark study that uses Bayesian inference for reliable prediction of molecular properties\cite{hwang2020comprehensive},
we used Monte-Carlo dropout (MC-DO)\cite{gal2016dropout} and stochastic weight averaging (SWA)\cite{maddox2019simple} to develop regression and classification models, respectively.

\subsection{Code availability}
We release our code in \url{https://github.com/SeongokRyu/gnn_solubility} to clarify implementations and reproducibility of experimental results.

\section{Results and discussion}

\subsection{Attention-based readout and Bayesian learning leads to more accurate results}
 
\begin{figure}
    \centering
    \includegraphics[width=0.98\textwidth,trim={0cm 0 0cm 0},clip]{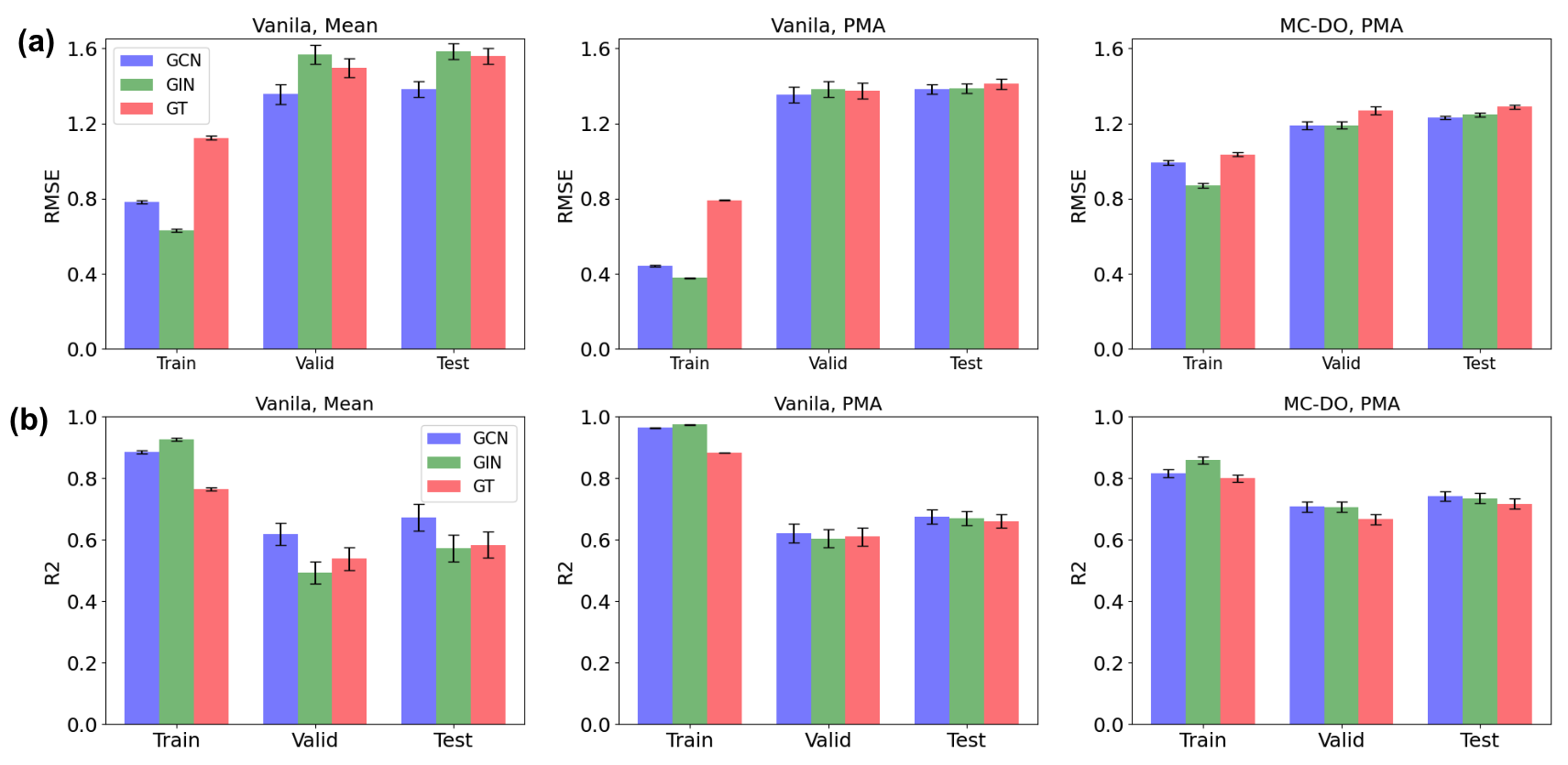}
    \caption{Prediction performance of regression models in terms of (a) root-mean-square-error (RMSE) and (b) coefficient of determination ($R^2$).}
    \label{fig:RMSE_R2}
\end{figure}

Firstly, we compare the performance of solubility regression models built on various model architectures and Bayesian learning algorithm in order to seek the best model specification. 
We investigate three node feature updates (i.e., GCN, GIN and GT) and two graph feature updates (i.e., mean and PMA). 
Also, we test the usefulness of approximate Bayesian learning with MC-DO.

In Figure \ref{fig:RMSE_R2}, we show the performances in terms of root-mean-square-error (RMSE) and coefficient of determination ($R^2$). 
Contrary to previous researches, which showed that using self-attention (i.e. GT) or edge features (i.e. GIN) improve prediction power, 
the simplest node update (i.e. GCN and GT) shows the best predictive performance among three node updates. 
On the other hand, regardless of node updating methods, the attention-based readout (i.e. PMA) consistently outperforms the mean readout. 
The results imply that fine-tuning the readout layer of GNN architectures is more helpful than tuning the node update layers to improve the predictive power, at least in our solubility prediction task.

Next, by comparing the point-estimation of model parameters (labeled as `Vanila') and Bayesian model (labeled as `MC-DO'), 
we confirm that adopting approximate Bayesian learning significantly improves the predictive performance. 
Although using MC-DO gives us higher RMSE and lower $R^2$ on the train set, it gives us lower RMSE and $R^2$ on the validation and test sets,
and thereby reduces the generalization gap between the train set and validation/test sets.

\begin{figure}
    \centering
    \includegraphics[width=0.48\textwidth,trim={0cm 0 0cm 0},clip]{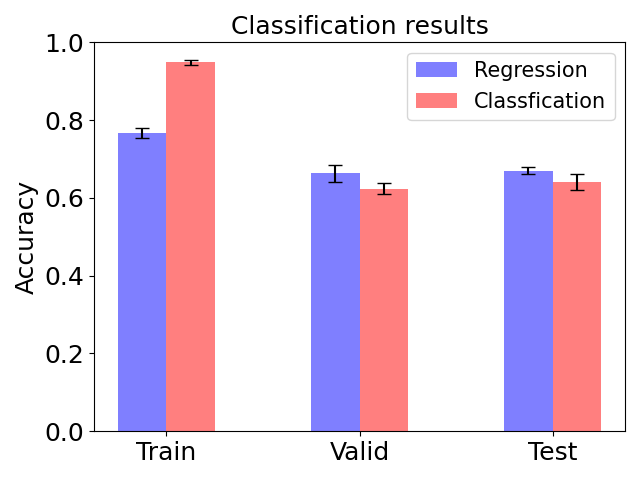}
    \caption{Classification accuracy of the regression and classification models. We note that the prediction results from the regression model were converted to the categorical values and evaluated in terms of classification accuracy.}
    \label{fig:classification}
\end{figure}

Since the solubility indicator value given by the AqSolDB dataset can be categorized into four classes, as described in the method section, we trained the classification model and evaluated the classification performance of the best regression model by converting true labels and predictive labels to categorical values.
Figure \ref{fig:classification} demonstrates the accuracy of both the regression and the classification models.
Though the regression model underperformed in the train set, opposite results were obtained in the validation and test sets. 
We conjecture that the training setting of regression tasks might be more information-rich than that of classification tasks.
For example, indicator values -1.9 and -0.1 are both categorized into the `soluble' group, but the former may show smaller solubility than the latter. 
On the other hand, categorical values are identical for both values, provides the same information to the model during the training stage. 
Therefore, continuous indicator values provide more information than categorical values, resulting in better generalization results.

Based on aforementioned results, we chose the regression model built on the GCN node update, the PMA readout, and the MC-DO Bayesian inference as the baseline architecture for further investigation.

\subsection{More reliable predictions can be prioritized with predictive uncertainty}

\begin{figure}
    \centering
    \includegraphics[width=0.98\textwidth,trim={0cm 0 0cm 0},clip]{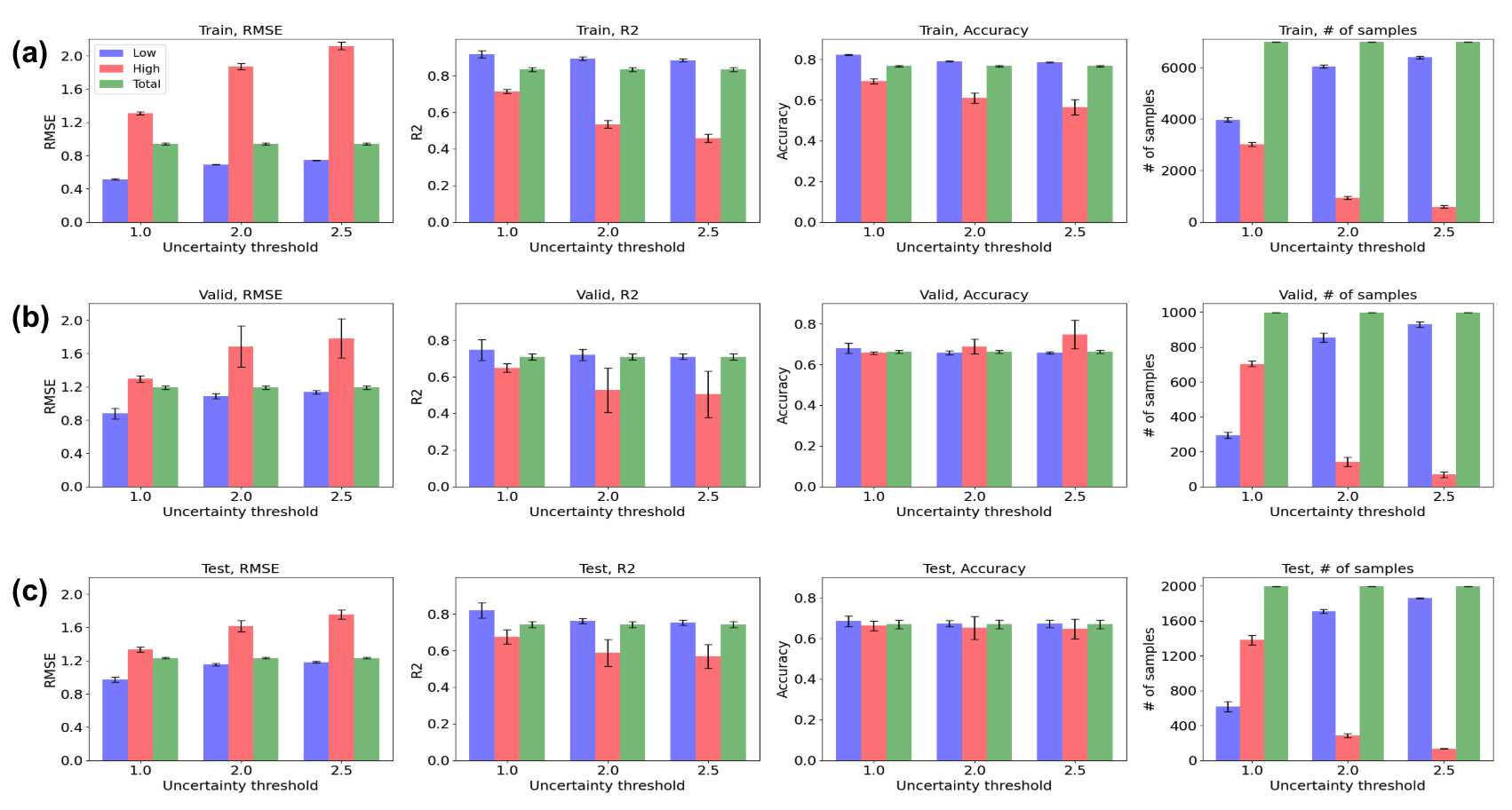}
    \caption{Prediction performance of the best regression models for (a) train, (b) validation and (c) test sets. We report the prediction performance in terms of root-mean-square-error (RMSE), coefficient of determination ($R^2$) and classification accuracy. The right-most panel shows the number of samples in the low and high uncertainty groups and the total number of samples.}
    \label{fig:uncertainty_aware}
\end{figure}

One of the biggest advantages of Bayesian learning is utilizing uncertainty to distinguish reliable and unreliable results predictions.
For example, \citeauthor{ryu2019bayesian}\cite{ryu2019bayesian} demonstrated that mis-labeled noisy data can be detected by high aleatoric (data-driven) uncertainty and elucidated the source of those errors.
Inspired from such applications, we investigated the reliability of our solubility prediction model. 

As shown in Figure \ref{fig:uncertainty_aware}, we categorize the prediction results into high and low uncertainty groups and report the predictive performances in terms of RMSE, $R^2$ and classification accuracy. 
The x-axis display the uncertainty thresholds, with which we divide the uncertainty groups
As expected, the lower the uncertainty groups, the better their predictive performances in terms of lower RMSE, higher $R^2$ and higher classification accuracy.
One exceptional was higher classification accuracy for higher uncertainty group in the validation set, which possibly due to extremely small number of samples used for validation, as is shown in the right-most panel of Figure \ref{fig:uncertainty_aware}(b).

\subsection{Visualization and interpretation of prediction results with attention weights}

Understanding how black-box machine learning models perform a given task is crucial point for reliability.
Therefore, we interpret solubility prediction results on FDA-approved drug compounds by investigating the changes in predictive solubility according to structural modification and visualizing of the attention weights computed by the PMA readout.

\begin{figure}
    \centering
    \includegraphics[width=0.95\textwidth,trim={0cm 0 0cm 0},clip]{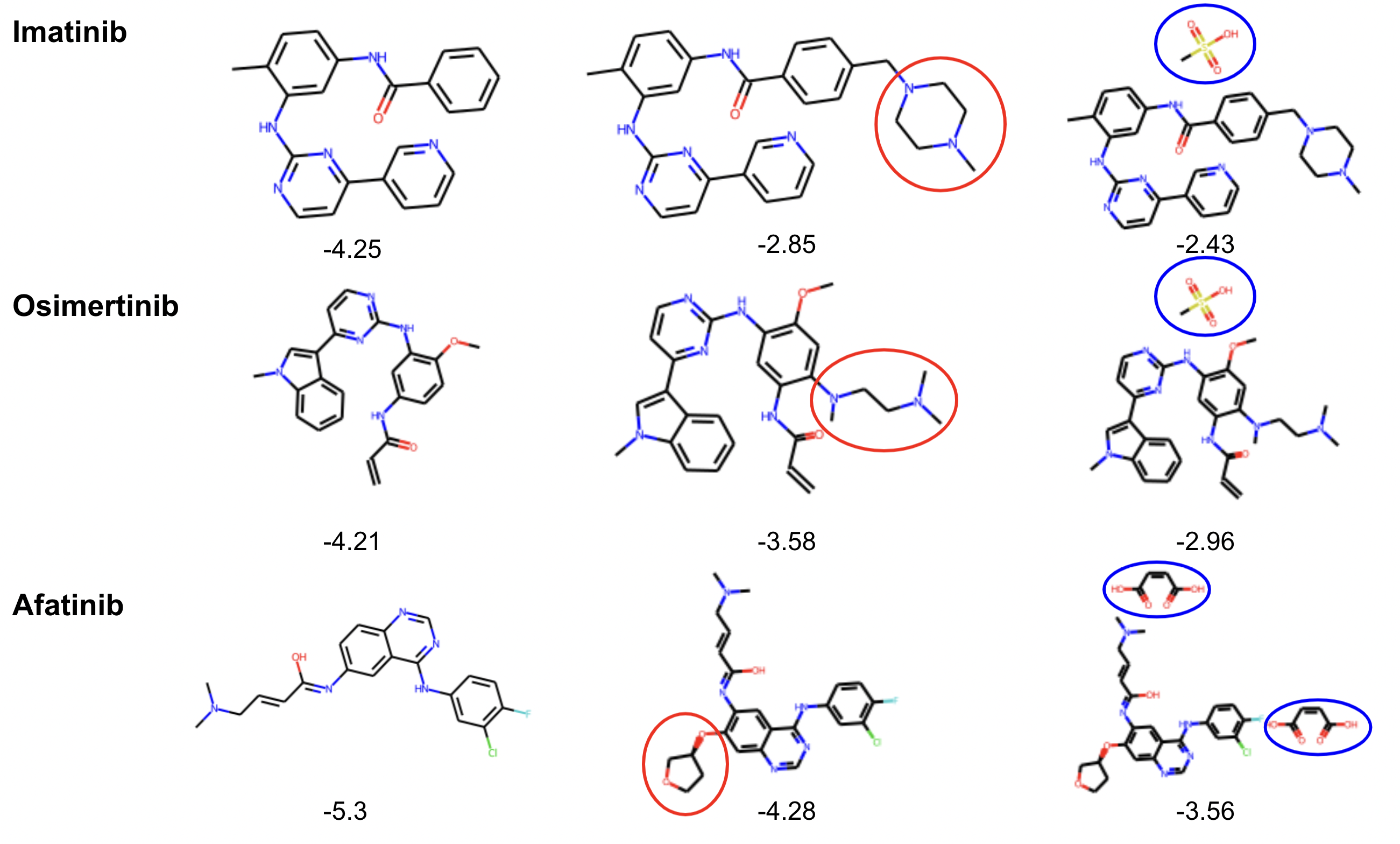}
    \caption{FDA approved drugs -- imatinib, osimertinib, and afatinib -- and their predictive solubility. We show the neutral form of the drugs (middle), their salt form (right) and the detached form of the solubility increasing group (left).}
    \label{fig:lead_opt_example}
\end{figure}

In Figure \ref{fig:lead_opt_example}, we show three FDA-approved drugs (middle), i.e. imatinib, osimertinib and afatinib, their salt-form (right), and the compounds before structural modification (left). 
Predictive solubility of each compound is noted below. 
We highlight the attached substructures, which are commonly used to enhance solubility and bio-availability, to the left-most compounds in red circles (middle) and anionic salts in blue circles (right). 
Consistent with medicinal chemistry knowledge, such structural modifications increase predictive solubility. 
For example, \textit{N}-methyl piperazine group is well-known for increasing bioavailibity and have been used to optimize lead compounds; 
our model predicted that it significantly increases solubility through structural modification of the imatinib. 
Similar results were observed in modifications of osimertinib and afatinib with hydrophilic motifs highlighted in red circles, respectively. 
Not so surprisingly, input graphs represented with salt forms also showed increased predictive solubility compared to the neutral forms.
Albeit we cannot ensure that experimental values of those compounds, we propose that our model can be used for one part of lead optimization steps -- improve solubility by structure modification.

\begin{figure}
    \centering
    \includegraphics[width=0.98\textwidth,trim={0cm 0 0cm 0},clip]{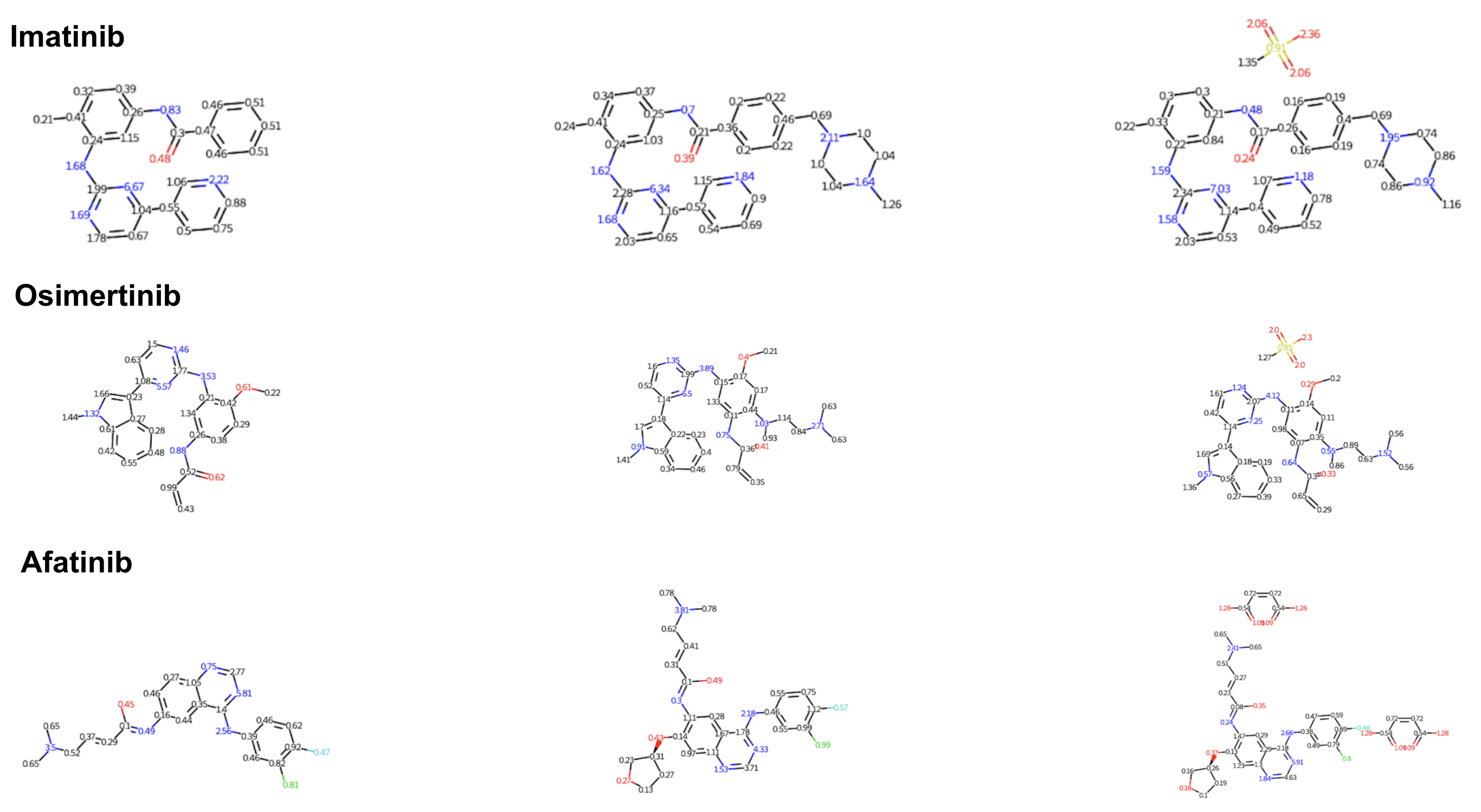}
    \caption{The attention weights of the atoms of three FDA approved drugs -- imatinib, osimertinib, and afatinib. We show the neutral form of the drugs (middle), their salt form (right) and the detached form of the solubility increasing group (left).}
    \label{fig:attention}
\end{figure}

To interpret which atoms are considered by our prediction model to be important, in Figure \ref{fig:attention}, 
we visualize the attention weights $\{\bar{\alpha}_v\}_{v \in V}$ computed by the PMA readout.
The attention weight of the $v$-th node is computed as follows:
\begin{equation}
    \bar{\alpha}_v = \frac{N_{at}}{T \cdot H} \sum_{t=1}^{T} \sum_{h=1}^{H} \alpha_{vht}, 
\end{equation}
where $\alpha_{vht}$ is the attention weight for the $v$-th node from the $h$-th attention head at the $t$-th MC-sampling step, $T$ is the number of MC-samplings for MC-DO, $H$ is the number heads for multi-head attention, and $N_{at}$ is the number of atoms for a given molecule.
Since the softmax activation normalizes attention weights to one, we scale attention weights by multiplying the number of atoms to prevent attention weights to be small for large molecules.

For all three examples, we observe higher attention weights for hetero-aromatic rings than benzene rings, which is consistent to the fact that hetero-aromatic rings are more polar and hydrophilic than benzene rings. 
Also, consistent to common medicinal chemistry knowledge and solubility prediction results, our model gives high attention weights to newly attached motifs, which are highlighted with red circles in the middle panel of Figure \ref{fig:lead_opt_example}. 
WE observe similar results for anionic salts, which are highlighted in blue circles in the right-most panel of Figure \ref{fig:lead_opt_example}.

\section{Conclusion}

Here, we present a solubility prediction model based on graph neural networks, self-attention and Bayesian inference.
While previous studies focused on advancing node update parts, we find that using self-attention in the readout part is more effective in improving prediction accuracy. 
This may be because, our attention weights are interpreted as importance of atoms for a given prediction task.
The experimental results demonstrate that such interpretation is valid, solubility-increasing substructures showed high attention weights.
Also, Bayesian inference via approximate MC-dropout enables us to quantify predictive uncertainty and differentiate between more and less accurate results.
Overall, our prediction results help us choose virtually screened candidates for experimental assays by analyzing uncertainty and attention weights.

We would like to assert that our prediction framework can be extended to any ligand-based (single-instance) prediction task.
Recently, Therapeutic Data Commons, the consortium to benchmark molecular machine learning,\cite{huang2021therapeutics} has provided a variety of well-curated datasets including prediction tasks, with both single-instance and multi-instances, and generative tasks. 
Among the single-instance tasks, prediction of ADME and toxicity are as important as solubility prediction to reduce failed trials in drug developments.
Since provided datasets are also data-deficient, prediction models may be vulnerable to over-fitting issue. 
For those tasks, we highlight that assessment of reliability and interpretability is also essential for robust decision making and communication with experimentalists.\cite{schwaller2019molecular, soleimany2021evidential} 

In addition, we believe that our solubility prediction model can be extended to other machine learning applications, such as active learning and generative reinforcement learning.
Predictive uncertainty indicates which data points require additional labeling in supervised learning and exploration in (generative-) reinforcement learning.\cite{gal2017deep}
Such application have gained increasing attention to maximize performance with limited computational or experimental costs. \cite{graff2021accelerating, soleimany2021evidential, gentile2020deep, yang2021hit}
Studying how increases in training/explored data points change predictive uncertainty and attention weights enables us to draw better understanding of molecular machine learning with deep neural networks.
We leave aforementioned discussion as possible future research direction.

\section*{Acknowledgements}
We would like to appreciate Juwon Hong and Chaok Seok for proofread of manuscripts.
This work was supported by Galux Inc. (No. Galux-20210001).

\section*{Author contributions}
Seongok Ryu conceived the idea and performed implementation and experiments. 
Seongok Ryu and Sumin Lee analyzed the results and wrote the manuscript together.

\section*{Conflicts of interest}
The authors declare no competing financial interests.

\begin{tocentry}

\begin{center}
    \centering
    \includegraphics[width=0.85\textwidth,trim={0cm 0 0cm 0},clip]{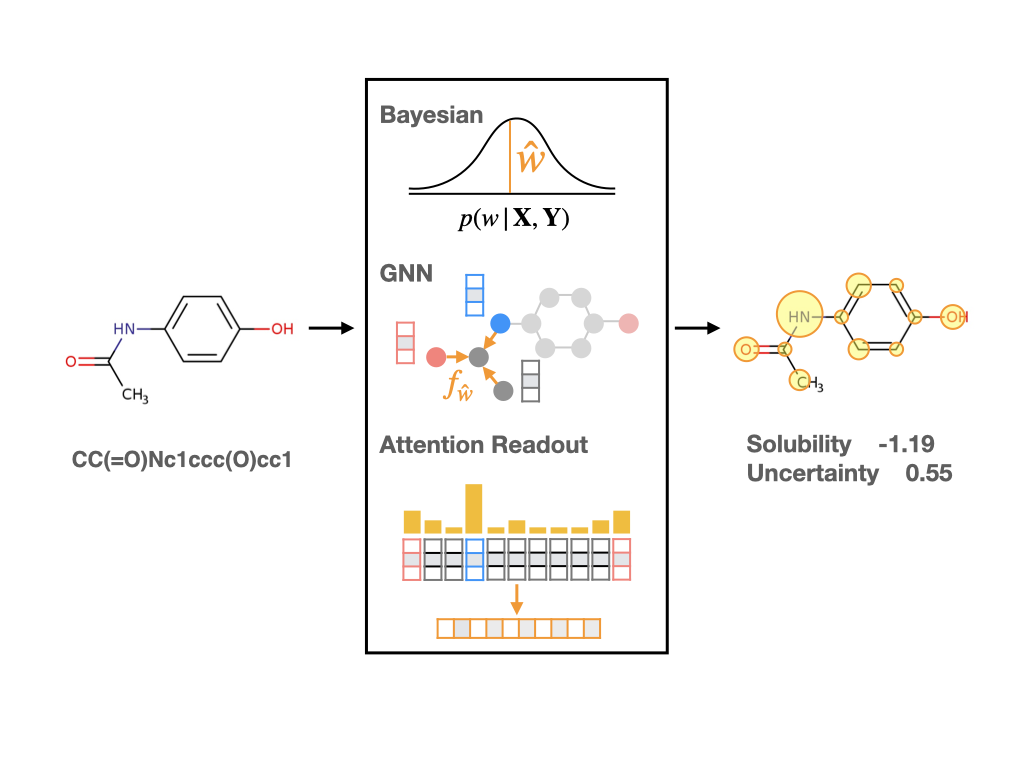}
\end{center}

\end{tocentry}

\bibliography{achemso-demo}

\providecommand{\latin}[1]{#1}
\makeatletter
\providecommand{\doi}
  {\begingroup\let\do\@makeother\dospecials
  \catcode`\{=1 \catcode`\}=2 \doi@aux}
\providecommand{\doi@aux}[1]{\endgroup\texttt{#1}}
\makeatother
\providecommand*\mcitethebibliography{\thebibliography}
\csname @ifundefined\endcsname{endmcitethebibliography}
  {\let\endmcitethebibliography\endthebibliography}{}
\begin{mcitethebibliography}{33}
\providecommand*\natexlab[1]{#1}
\providecommand*\mciteSetBstSublistMode[1]{}
\providecommand*\mciteSetBstMaxWidthForm[2]{}
\providecommand*\mciteBstWouldAddEndPuncttrue
  {\def\EndOfBibitem{\unskip.}}
\providecommand*\mciteBstWouldAddEndPunctfalse
  {\let\EndOfBibitem\relax}
\providecommand*\mciteSetBstMidEndSepPunct[3]{}
\providecommand*\mciteSetBstSublistLabelBeginEnd[3]{}
\providecommand*\EndOfBibitem{}
\mciteSetBstSublistMode{f}
\mciteSetBstMaxWidthForm{subitem}{(\alph{mcitesubitemcount})}
\mciteSetBstSublistLabelBeginEnd
  {\mcitemaxwidthsubitemform\space}
  {\relax}
  {\relax}

\bibitem[Landrum(2013)]{landrum2013rdkit}
Landrum,~G. RDKit: A software suite for cheminformatics, computational
  chemistry, and predictive modeling. 2013\relax
\mciteBstWouldAddEndPuncttrue
\mciteSetBstMidEndSepPunct{\mcitedefaultmidpunct}
{\mcitedefaultendpunct}{\mcitedefaultseppunct}\relax
\EndOfBibitem
\bibitem[Sorkun \latin{et~al.}(2019)Sorkun, Khetan, and Er]{sorkun2019aqsoldb}
Sorkun,~M.~C.; Khetan,~A.; Er,~S. AqSolDB, a curated reference set of aqueous
  solubility and 2D descriptors for a diverse set of compounds.
  \emph{Scientific data} \textbf{2019}, \emph{6}, 1--8\relax
\mciteBstWouldAddEndPuncttrue
\mciteSetBstMidEndSepPunct{\mcitedefaultmidpunct}
{\mcitedefaultendpunct}{\mcitedefaultseppunct}\relax
\EndOfBibitem
\bibitem[Klamt \latin{et~al.}(2002)Klamt, Eckert, Hornig, Beck, and
  B{\"u}rger]{klamt2002prediction}
Klamt,~A.; Eckert,~F.; Hornig,~M.; Beck,~M.~E.; B{\"u}rger,~T. Prediction of
  aqueous solubility of drugs and pesticides with COSMO-RS. \emph{Journal of
  computational chemistry} \textbf{2002}, \emph{23}, 275--281\relax
\mciteBstWouldAddEndPuncttrue
\mciteSetBstMidEndSepPunct{\mcitedefaultmidpunct}
{\mcitedefaultendpunct}{\mcitedefaultseppunct}\relax
\EndOfBibitem
\bibitem[Bjelobrk \latin{et~al.}(2021)Bjelobrk, Mendels, Karmakar, Parrinello,
  and Mazzotti]{bjelobrk2021solubility}
Bjelobrk,~Z.; Mendels,~D.; Karmakar,~T.; Parrinello,~M.; Mazzotti,~M.
  Solubility prediction of organic molecules with molecular dynamics
  simulations. \emph{Crystal Growth \& Design} \textbf{2021}, \emph{21},
  5198--5205\relax
\mciteBstWouldAddEndPuncttrue
\mciteSetBstMidEndSepPunct{\mcitedefaultmidpunct}
{\mcitedefaultendpunct}{\mcitedefaultseppunct}\relax
\EndOfBibitem
\bibitem[Sorkun \latin{et~al.}(2021)Sorkun, Koelman, and Er]{sorkun2021pushing}
Sorkun,~M.~C.; Koelman,~J. V.~A.; Er,~S. Pushing the limits of solubility
  prediction via quality-oriented data selection. \emph{Iscience}
  \textbf{2021}, \emph{24}, 101961\relax
\mciteBstWouldAddEndPuncttrue
\mciteSetBstMidEndSepPunct{\mcitedefaultmidpunct}
{\mcitedefaultendpunct}{\mcitedefaultseppunct}\relax
\EndOfBibitem
\bibitem[Lovri{\'c} \latin{et~al.}(2021)Lovri{\'c}, Pavlovi{\'c}, {\v{Z}}uvela,
  Spataru, Lu{\v{c}}i{\'c}, Kern, and Wong]{lovric2021machine}
Lovri{\'c},~M.; Pavlovi{\'c},~K.; {\v{Z}}uvela,~P.; Spataru,~A.;
  Lu{\v{c}}i{\'c},~B.; Kern,~R.; Wong,~M.~W. Machine learning in prediction of
  intrinsic aqueous solubility of drug-like compounds: Generalization,
  complexity, or predictive ability? \emph{Journal of Chemometrics}
  \textbf{2021}, \emph{35}, e3349\relax
\mciteBstWouldAddEndPuncttrue
\mciteSetBstMidEndSepPunct{\mcitedefaultmidpunct}
{\mcitedefaultendpunct}{\mcitedefaultseppunct}\relax
\EndOfBibitem
\bibitem[Llin{\`a}s \latin{et~al.}(2008)Llin{\`a}s, Glen, and
  Goodman]{llinas2008solubility}
Llin{\`a}s,~A.; Glen,~R.~C.; Goodman,~J.~M. Solubility challenge: can you
  predict solubilities of 32 molecules using a database of 100 reliable
  measurements? \emph{Journal of chemical information and modeling}
  \textbf{2008}, \emph{48}, 1289--1303\relax
\mciteBstWouldAddEndPuncttrue
\mciteSetBstMidEndSepPunct{\mcitedefaultmidpunct}
{\mcitedefaultendpunct}{\mcitedefaultseppunct}\relax
\EndOfBibitem
\bibitem[Hopfinger \latin{et~al.}(2009)Hopfinger, Esposito, Llinas, Glen, and
  Goodman]{hopfinger2009findings}
Hopfinger,~A.~J.; Esposito,~E.~X.; Llinas,~A.; Glen,~R.~C.; Goodman,~J.~M.
  Findings of the challenge to predict aqueous solubility. \emph{Journal of
  chemical information and modeling} \textbf{2009}, \emph{49}, 1--5\relax
\mciteBstWouldAddEndPuncttrue
\mciteSetBstMidEndSepPunct{\mcitedefaultmidpunct}
{\mcitedefaultendpunct}{\mcitedefaultseppunct}\relax
\EndOfBibitem
\bibitem[Palmer and Mitchell(2014)Palmer, and Mitchell]{palmer2014experimental}
Palmer,~D.~S.; Mitchell,~J.~B. Is experimental data quality the limiting factor
  in predicting the aqueous solubility of druglike molecules? \emph{Molecular
  Pharmaceutics} \textbf{2014}, \emph{11}, 2962--2972\relax
\mciteBstWouldAddEndPuncttrue
\mciteSetBstMidEndSepPunct{\mcitedefaultmidpunct}
{\mcitedefaultendpunct}{\mcitedefaultseppunct}\relax
\EndOfBibitem
\bibitem[Llinas \latin{et~al.}(2020)Llinas, Oprisiu, and
  Avdeef]{llinas2020findings}
Llinas,~A.; Oprisiu,~I.; Avdeef,~A. Findings of the second challenge to predict
  aqueous solubility. \emph{Journal of chemical information and modeling}
  \textbf{2020}, \emph{60}, 4791--4803\relax
\mciteBstWouldAddEndPuncttrue
\mciteSetBstMidEndSepPunct{\mcitedefaultmidpunct}
{\mcitedefaultendpunct}{\mcitedefaultseppunct}\relax
\EndOfBibitem
\bibitem[Ryu \latin{et~al.}(2019)Ryu, Kwon, and Kim]{ryu2019bayesian}
Ryu,~S.; Kwon,~Y.; Kim,~W.~Y. A Bayesian graph convolutional network for
  reliable prediction of molecular properties with uncertainty quantification.
  \emph{Chemical science} \textbf{2019}, \emph{10}, 8438--8446\relax
\mciteBstWouldAddEndPuncttrue
\mciteSetBstMidEndSepPunct{\mcitedefaultmidpunct}
{\mcitedefaultendpunct}{\mcitedefaultseppunct}\relax
\EndOfBibitem
\bibitem[Schwaller \latin{et~al.}(2019)Schwaller, Laino, Gaudin, Bolgar,
  Hunter, Bekas, and Lee]{schwaller2019molecular}
Schwaller,~P.; Laino,~T.; Gaudin,~T.; Bolgar,~P.; Hunter,~C.~A.; Bekas,~C.;
  Lee,~A.~A. Molecular transformer: a model for uncertainty-calibrated chemical
  reaction prediction. \emph{ACS central science} \textbf{2019}, \emph{5},
  1572--1583\relax
\mciteBstWouldAddEndPuncttrue
\mciteSetBstMidEndSepPunct{\mcitedefaultmidpunct}
{\mcitedefaultendpunct}{\mcitedefaultseppunct}\relax
\EndOfBibitem
\bibitem[Soleimany \latin{et~al.}(2021)Soleimany, Amini, Goldman, Rus, Bhatia,
  and Coley]{soleimany2021evidential}
Soleimany,~A.~P.; Amini,~A.; Goldman,~S.; Rus,~D.; Bhatia,~S.~N.; Coley,~C.~W.
  Evidential deep learning for guided molecular property prediction and
  discovery. \emph{ACS central science} \textbf{2021}, \emph{7},
  1356--1367\relax
\mciteBstWouldAddEndPuncttrue
\mciteSetBstMidEndSepPunct{\mcitedefaultmidpunct}
{\mcitedefaultendpunct}{\mcitedefaultseppunct}\relax
\EndOfBibitem
\bibitem[Graff \latin{et~al.}(2021)Graff, Shakhnovich, and
  Coley]{graff2021accelerating}
Graff,~D.~E.; Shakhnovich,~E.~I.; Coley,~C.~W. Accelerating high-throughput
  virtual screening through molecular pool-based active learning.
  \emph{Chemical science} \textbf{2021}, \emph{12}, 7866--7881\relax
\mciteBstWouldAddEndPuncttrue
\mciteSetBstMidEndSepPunct{\mcitedefaultmidpunct}
{\mcitedefaultendpunct}{\mcitedefaultseppunct}\relax
\EndOfBibitem
\bibitem[Harren \latin{et~al.}(2022)Harren, Matter, Hessler, Rarey, and
  Grebner]{harren2022interpretation}
Harren,~T.; Matter,~H.; Hessler,~G.; Rarey,~M.; Grebner,~C. Interpretation of
  Structure--Activity Relationships in Real-World Drug Design Data Sets Using
  Explainable Artificial Intelligence. \emph{Journal of Chemical Information
  and Modeling} \textbf{2022}, \relax
\mciteBstWouldAddEndPunctfalse
\mciteSetBstMidEndSepPunct{\mcitedefaultmidpunct}
{}{\mcitedefaultseppunct}\relax
\EndOfBibitem
\bibitem[Vaswani \latin{et~al.}(2017)Vaswani, Shazeer, Parmar, Uszkoreit,
  Jones, Gomez, Kaiser, and Polosukhin]{vaswani2017attention}
Vaswani,~A.; Shazeer,~N.; Parmar,~N.; Uszkoreit,~J.; Jones,~L.; Gomez,~A.~N.;
  Kaiser,~{\L}.; Polosukhin,~I. Attention is all you need. Advances in neural
  information processing systems. 2017; pp 5998--6008\relax
\mciteBstWouldAddEndPuncttrue
\mciteSetBstMidEndSepPunct{\mcitedefaultmidpunct}
{\mcitedefaultendpunct}{\mcitedefaultseppunct}\relax
\EndOfBibitem
\bibitem[Huang \latin{et~al.}(2021)Huang, Fu, Gao, Zhao, Roohani, Leskovec,
  Coley, Xiao, Sun, and Zitnik]{huang2021therapeutics}
Huang,~K.; Fu,~T.; Gao,~W.; Zhao,~Y.; Roohani,~Y.; Leskovec,~J.; Coley,~C.~W.;
  Xiao,~C.; Sun,~J.; Zitnik,~M. Therapeutics data Commons: machine learning
  datasets and tasks for therapeutics. \emph{arXiv preprint arXiv:2102.09548}
  \textbf{2021}, \relax
\mciteBstWouldAddEndPunctfalse
\mciteSetBstMidEndSepPunct{\mcitedefaultmidpunct}
{}{\mcitedefaultseppunct}\relax
\EndOfBibitem
\bibitem[Wu \latin{et~al.}(2018)Wu, Ramsundar, Feinberg, Gomes, Geniesse,
  Pappu, Leswing, and Pande]{wu2018moleculenet}
Wu,~Z.; Ramsundar,~B.; Feinberg,~E.~N.; Gomes,~J.; Geniesse,~C.; Pappu,~A.~S.;
  Leswing,~K.; Pande,~V. MoleculeNet: a benchmark for molecular machine
  learning. \emph{Chemical science} \textbf{2018}, \emph{9}, 513--530\relax
\mciteBstWouldAddEndPuncttrue
\mciteSetBstMidEndSepPunct{\mcitedefaultmidpunct}
{\mcitedefaultendpunct}{\mcitedefaultseppunct}\relax
\EndOfBibitem
\bibitem[Hwang \latin{et~al.}(2020)Hwang, Yang, Kwon, Lee, Lee, Jo, Yoon, and
  Ryu]{hwang2020comprehensive}
Hwang,~D.; Yang,~S.; Kwon,~Y.; Lee,~K.~H.; Lee,~G.; Jo,~H.; Yoon,~S.; Ryu,~S.
  Comprehensive Study on Molecular Supervised Learning with Graph Neural
  Networks. \emph{Journal of Chemical Information and Modeling} \textbf{2020},
  \emph{60}, 5936--5945\relax
\mciteBstWouldAddEndPuncttrue
\mciteSetBstMidEndSepPunct{\mcitedefaultmidpunct}
{\mcitedefaultendpunct}{\mcitedefaultseppunct}\relax
\EndOfBibitem
\bibitem[Duvenaud \latin{et~al.}(2015)Duvenaud, Maclaurin, Iparraguirre,
  Bombarell, Hirzel, Aspuru-Guzik, and Adams]{duvenaud2015convolutional}
Duvenaud,~D.~K.; Maclaurin,~D.; Iparraguirre,~J.; Bombarell,~R.; Hirzel,~T.;
  Aspuru-Guzik,~A.; Adams,~R.~P. Convolutional networks on graphs for learning
  molecular fingerprints. \emph{Advances in neural information processing
  systems} \textbf{2015}, \emph{28}\relax
\mciteBstWouldAddEndPuncttrue
\mciteSetBstMidEndSepPunct{\mcitedefaultmidpunct}
{\mcitedefaultendpunct}{\mcitedefaultseppunct}\relax
\EndOfBibitem
\bibitem[Kipf and Welling(2016)Kipf, and Welling]{kipf2016semi}
Kipf,~T.~N.; Welling,~M. Semi-supervised classification with graph
  convolutional networks. \emph{arXiv preprint arXiv:1609.02907} \textbf{2016},
  \relax
\mciteBstWouldAddEndPunctfalse
\mciteSetBstMidEndSepPunct{\mcitedefaultmidpunct}
{}{\mcitedefaultseppunct}\relax
\EndOfBibitem
\bibitem[Ba \latin{et~al.}(2016)Ba, Kiros, and Hinton]{ba2016layer}
Ba,~J.~L.; Kiros,~J.~R.; Hinton,~G.~E. Layer normalization. \emph{arXiv
  preprint arXiv:1607.06450} \textbf{2016}, \relax
\mciteBstWouldAddEndPunctfalse
\mciteSetBstMidEndSepPunct{\mcitedefaultmidpunct}
{}{\mcitedefaultseppunct}\relax
\EndOfBibitem
\bibitem[Xu \latin{et~al.}(2018)Xu, Hu, Leskovec, and Jegelka]{xu2018powerful}
Xu,~K.; Hu,~W.; Leskovec,~J.; Jegelka,~S. How powerful are graph neural
  networks? \emph{arXiv preprint arXiv:1810.00826} \textbf{2018}, \relax
\mciteBstWouldAddEndPunctfalse
\mciteSetBstMidEndSepPunct{\mcitedefaultmidpunct}
{}{\mcitedefaultseppunct}\relax
\EndOfBibitem
\bibitem[Veli{\v{c}}kovi{\'c} \latin{et~al.}(2017)Veli{\v{c}}kovi{\'c},
  Cucurull, Casanova, Romero, Lio, and Bengio]{velivckovic2017graph}
Veli{\v{c}}kovi{\'c},~P.; Cucurull,~G.; Casanova,~A.; Romero,~A.; Lio,~P.;
  Bengio,~Y. Graph attention networks. \emph{arXiv preprint arXiv:1710.10903}
  \textbf{2017}, \relax
\mciteBstWouldAddEndPunctfalse
\mciteSetBstMidEndSepPunct{\mcitedefaultmidpunct}
{}{\mcitedefaultseppunct}\relax
\EndOfBibitem
\bibitem[Fey and Lenssen(2019)Fey, and Lenssen]{fey2019fast}
Fey,~M.; Lenssen,~J.~E. Fast graph representation learning with PyTorch
  Geometric. \emph{arXiv preprint arXiv:1903.02428} \textbf{2019}, \relax
\mciteBstWouldAddEndPunctfalse
\mciteSetBstMidEndSepPunct{\mcitedefaultmidpunct}
{}{\mcitedefaultseppunct}\relax
\EndOfBibitem
\bibitem[Dosovitskiy \latin{et~al.}(2020)Dosovitskiy, Beyer, Kolesnikov,
  Weissenborn, Zhai, Unterthiner, Dehghani, Minderer, Heigold, Gelly,
  \latin{et~al.} others]{dosovitskiy2020image}
Dosovitskiy,~A.; Beyer,~L.; Kolesnikov,~A.; Weissenborn,~D.; Zhai,~X.;
  Unterthiner,~T.; Dehghani,~M.; Minderer,~M.; Heigold,~G.; Gelly,~S.,
  \latin{et~al.}  An image is worth 16x16 words: Transformers for image
  recognition at scale. \emph{arXiv preprint arXiv:2010.11929} \textbf{2020},
  \relax
\mciteBstWouldAddEndPunctfalse
\mciteSetBstMidEndSepPunct{\mcitedefaultmidpunct}
{}{\mcitedefaultseppunct}\relax
\EndOfBibitem
\bibitem[Lee \latin{et~al.}(2019)Lee, Lee, Kim, Kosiorek, Choi, and
  Teh]{lee2019set}
Lee,~J.; Lee,~Y.; Kim,~J.; Kosiorek,~A.; Choi,~S.; Teh,~Y.~W. Set transformer:
  A framework for attention-based permutation-invariant neural networks.
  International Conference on Machine Learning. 2019; pp 3744--3753\relax
\mciteBstWouldAddEndPuncttrue
\mciteSetBstMidEndSepPunct{\mcitedefaultmidpunct}
{\mcitedefaultendpunct}{\mcitedefaultseppunct}\relax
\EndOfBibitem
\bibitem[Gal and Ghahramani(2016)Gal, and Ghahramani]{gal2016dropout}
Gal,~Y.; Ghahramani,~Z. Dropout as a bayesian approximation: Representing model
  uncertainty in deep learning. international conference on machine learning.
  2016; pp 1050--1059\relax
\mciteBstWouldAddEndPuncttrue
\mciteSetBstMidEndSepPunct{\mcitedefaultmidpunct}
{\mcitedefaultendpunct}{\mcitedefaultseppunct}\relax
\EndOfBibitem
\bibitem[Maddox \latin{et~al.}(2019)Maddox, Izmailov, Garipov, Vetrov, and
  Wilson]{maddox2019simple}
Maddox,~W.~J.; Izmailov,~P.; Garipov,~T.; Vetrov,~D.~P.; Wilson,~A.~G. A simple
  baseline for bayesian uncertainty in deep learning. \emph{Advances in Neural
  Information Processing Systems} \textbf{2019}, \emph{32}\relax
\mciteBstWouldAddEndPuncttrue
\mciteSetBstMidEndSepPunct{\mcitedefaultmidpunct}
{\mcitedefaultendpunct}{\mcitedefaultseppunct}\relax
\EndOfBibitem
\bibitem[Gal \latin{et~al.}(2017)Gal, Islam, and Ghahramani]{gal2017deep}
Gal,~Y.; Islam,~R.; Ghahramani,~Z. Deep bayesian active learning with image
  data. International Conference on Machine Learning. 2017; pp 1183--1192\relax
\mciteBstWouldAddEndPuncttrue
\mciteSetBstMidEndSepPunct{\mcitedefaultmidpunct}
{\mcitedefaultendpunct}{\mcitedefaultseppunct}\relax
\EndOfBibitem
\bibitem[Gentile \latin{et~al.}(2020)Gentile, Agrawal, Hsing, Ton, Ban,
  Norinder, Gleave, and Cherkasov]{gentile2020deep}
Gentile,~F.; Agrawal,~V.; Hsing,~M.; Ton,~A.-T.; Ban,~F.; Norinder,~U.;
  Gleave,~M.~E.; Cherkasov,~A. Deep docking: a deep learning platform for
  augmentation of structure based drug discovery. \emph{ACS central science}
  \textbf{2020}, \emph{6}, 939--949\relax
\mciteBstWouldAddEndPuncttrue
\mciteSetBstMidEndSepPunct{\mcitedefaultmidpunct}
{\mcitedefaultendpunct}{\mcitedefaultseppunct}\relax
\EndOfBibitem
\bibitem[Yang \latin{et~al.}(2021)Yang, Hwang, Lee, Ryu, and
  Hwang]{yang2021hit}
Yang,~S.; Hwang,~D.; Lee,~S.; Ryu,~S.; Hwang,~S.~J. Hit and Lead Discovery with
  Explorative RL and Fragment-based Molecule Generation. \emph{Advances in
  Neural Information Processing Systems} \textbf{2021}, \emph{34}\relax
\mciteBstWouldAddEndPuncttrue
\mciteSetBstMidEndSepPunct{\mcitedefaultmidpunct}
{\mcitedefaultendpunct}{\mcitedefaultseppunct}\relax
\EndOfBibitem
\end{mcitethebibliography}

\end{document}